# From fire whirls to blue whirls and combustion without pollution

Huahua Xiao[a,1], Michael J. Gollner[b], and Elaine S. Oran[a]

[a]Department of Aerospace Engineering, University of Maryland, College Park, Maryland 20742, USA.

[b]Department of Fire Protection Engineering, University of Maryland, College Park, Maryland 20742, USA.

[1]To whom correspondence should be addressed. Tel: 1-301-412-6534. Email: hhxiao@umd.edu.

**Abstract**

Fire whirls are powerful, spinning disasters for people and surroundings when they occur in large urban and wildland fires. While fire whirls have long been studied for fire safety applications, previous research has yet to harness their potential burning efficiency for enhanced combustion. This paper presents laboratory studies of fire whirls initiated as pool fires, but where the fuel sits on a water surface, seeding an idea of exploiting the high-efficiency of fire whirls for oil-spill remediation. We show the transition from a pool fire, to a fire whirl, and then to a *previously unobserved state, a "blue whirl."* A blue whirl is smaller, very stable, and burns completely blue in a hydrocarbon flame, indicating soot-free burning. The combination of fast mixing and the water-surface boundary creates the conditions leading to nearly soot-free combustion. With the worldwide need to reduce emissions from both wanted and unwanted combustion, discovery of this new state points to new pathways for highly efficient and low-emission energy production and fuel-spill cleanup. Because current methods to generate a stable vortex are difficult, we also propose that the blue whirl may serve as a research platform for fundamental studies of vortices in fluid mechanics.

**Significance**

The growing worldwide demand to reduce emissions from combustion calls for development of alternative technologies of high-efficiency and low-emission combustion. While fire whirls are known for their intense and disastrous threat to life and surrounding environments, their high combustion efficiency implies a great potential for highly efficient and low-emission combustion, which has not yet been exploited. In studying fire whirls over water for oil-spill cleanup, we discovered a beautiful, swirling flame phenomenon, a *"blue whirl,"* which evolves from a fire whirl and burns completely with nearly soot-free combustion. Understanding and control of the blue whirl and its predecessor, the fire whirl, will suggest new ways for oil-spill remediation, clean combustion, energy production, and fluid mechanics research.

**Introduction**

Fire tornadoes, fire devils, and fire twisters are popular and terrifying common names for fire whirls. These intense swirling fires arise spontaneously with the "right" combination of wind and fire. On the large scale, they appear very similar to atmospheric phenomena such as tornadoes and dust devils (1-4). Fire whirls form in large urban and wildland fires when winds interact with obstacles or natural features in the terrain (2, 4-6) and produce large vortices that intensify as they interact with a local fire. When fire whirls arise naturally in large fires, they present a strong, essentially uncontrollable threat to life, property and surrounding environments. Due to the strong vertical winds they generate, they can lift and toss burning debris, which can then travel kilometers to spread the fire (5, 7).

A *pool fire* is a diffusion flame that burns above a horizontal pool of vaporizing hydrocarbon fuel. Pool fires can occur on any flat surface on which fuel is spread, including *in situ* burning of oil spills (8). *Fire whirls* can evolve from relatively small fires under proper wind or topographic conditions (1, 2, 5-7, 9). Smaller-scale laboratory experiments have shown that a relatively quiescent pool fire may transform into a fire whirl (2, 7, 10-14), and that the temperature and burning efficiency are higher than those of the initial fire (2, 10, 15). This transition occurs through a pattern of events in which the fire first leans to one side, begins to rotate, and then stretches upward (lengthens) before eventually becoming a fire whirl (7, 10, 14). Occasionally, extremely large and violent fire whirls appear when *in situ* burning is used for oil-spill remediation (16). They are noted when the otherwise black, sooty smoke turns nearly white. Higher-temperatures and thus cleaner burning than pool fires is a characteristic of fire whirls.

Flame heights from wildfires, pool fires and fire whirls range from centimeters to kilometers (1, 12, 14, 17, 18), with smaller experiments performed in laboratories so that their basic properties may be understood (2,4,6,7,13,19,20). For the purpose of improving fuel-spill remediation, we began a study of swirling flames ignited and burning on water, as opposed to the usual solid ground. First, we found that a fire whirl on land burns fuel more efficiently (faster and cleaner) than a pool fire, and second, that a fire whirl on water burns fuel much more efficiently than a fire whirl on a solid surface (21).

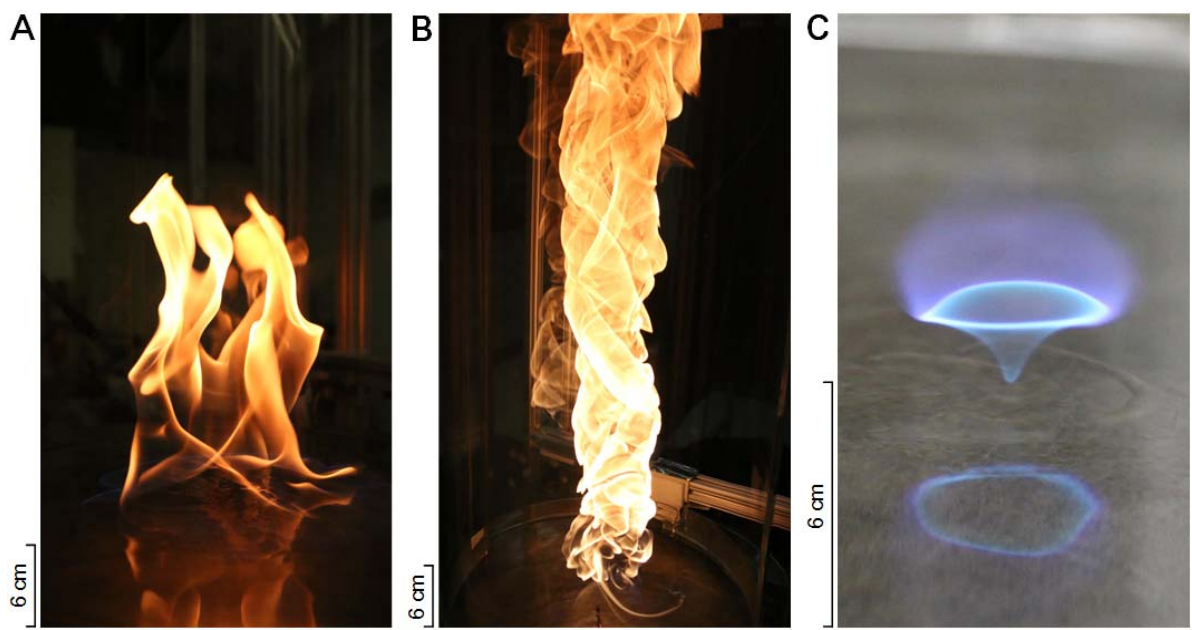

**Figure 1.** Evolution from a pool fire to a blue whirl over water in a swirl generator. (*A*) Pool fire forms following ignition. (*B*) Canonical fire whirl develops subsequent to the pool fire. (*C*) Newly-discovered laminar blue whirl evolves from the yellow fire whirl.

**Blue fire whirl over water**
A major, obvious difference between this and prior studies of pool fires and fires whirls on solid surfaces is the physical complexity of the boundary layer between the water, evaporating fuel, and flame. Figure 1 shows the three states of burning that we observe on water. Figures 1*A* and 1*B* are the usual pool fire and fire whirl, respectively, created here when winds enter the chamber tangentially. Figure 1*C* is the new flame structure discovered, the blue whirl, which evolves from the yellow fire whirl and is shown and discussed here for the first time.

The experimental setup consists of two half-cylinders and a cylindrical stainless steel pan full of water. A liquid fuel, n-heptane, is poured on the surface of quiescent water at the center of the pan and is then ignited by a small igniter. The two quartz half-cylinders are suspended over the pan. Offsetting the half cylinders creates two vertical slits that allow air to be entrained tangentially to the flame region, a method often used to create fire whirls for laboratory study (19). In order to sustain the blue whirl for further observation and study, we introduced a small copper tube under the water to pump heptane to the center region of the water surface at a fixed flow rate. When the fuel injection rate was set between 0.8 and 1.2 ml/min, and the size of the vertical gaps was 1.8-3.0 cm, the blue whirl could be sustained as long as fuel was supplied (the longest time sustained being just under 8 minutes).

After ignition, a small, chaotic pool fire forms (Fig. 1*A*). As cold entrained air is drawn into the chamber, the fire creates a strong vertical flow. Initially, the pool fire tilts and meanders (Movie S1), as reported previously (7, 10, 14), but then a canonical fire whirl, over 60 cm high, forms at the center of the apparatus (Fig. 1*B*). As with fire whirls on solid surfaces, this fire whirl on water burns much more vigorously than the initial pool fire, vortex motions are strong, the fire whirl is taller, and the temperature is higher. Then, *unexpectedly*, *the fire whirl continues to evolve to a new fire structure* (Fig. 1*C*): *a small, intensely whirling blue flame* (Movies S1 and S2).

Whereas the pool fire and the fire whirl are turbulent, the blue whirl shows no visible or aural signs of turbulence. *A stable blue whirl is very quiet*. The rotation was generally clockwise, consistent with the direction of air inflow from the gaps in the apparatus. Throughout its lifetime, the blue whirl generally revolves clockwise in the center region of the container at an angular speed within 6.3 rad/s.

**Structure of the blue whirl**
Figure 2 shows a front view of the structure of a blue whirl. This structure consists of two main regions: the bright blue spinning flame at the base and a faint conical violet flame sitting above and (possibly) partly in the central cup. The lower blue region appears more stable, approximately 2 cm high, while the height of the visible violet region varies from 2 to 6 cm. Thus the total height of the blue whirl was 4-8 cm, much shorter than the yellow fire whirl (over 60 cm) or pool fire (typically around 25 cm for this system). Near the fuel surface, the blue whirl tapers to a small rounded bottom. Above the bottom point, the blue whirl spreads out as it moves upward and resembles a spinning top. There is often a gap between the water and the bottom of the blue whirl. The top of the blue whirl ends sharply, although there is a secondary hazy violet flame above it. Both the height and diameter of the blue whirl decreases smoothly as it finally begins to die when all of the fuel is consumed (Movie S1).

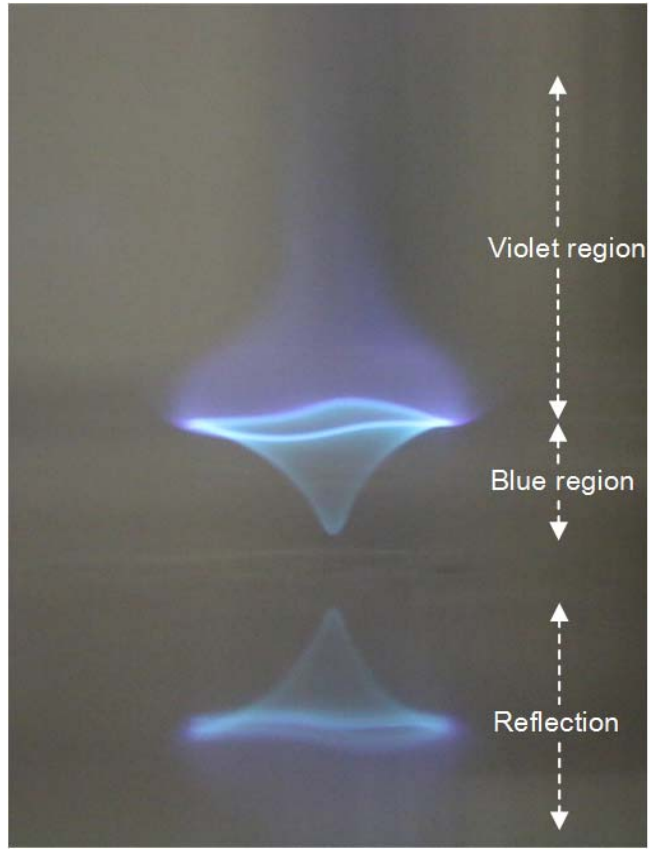

**Figure 2.** Front view of structure of a blue whirl generated over water. The structure mainly consist of a blue region and a violet region. A reflection of the blue whirl in the water can be seen in the lower part of the image.

**Transitions between yellow and blue whirls**

Once the yellow fire whirl has formed and burned for a time, it evolves into a transitional state with a diameter and height that fluctuate, but generally decrease (Movie S1). Eventually it forms a smaller "transitional whirl," which appears unstable and "dances around." There is a blue section at the base and a yellow flame that seems to grow out of this base, so that the blue whirl looks like a cup holding a yellow flame. A photograph of this transition state is shown in Fig. 3. After this, the yellow flame dies out leaving only a blue whirl.

During the lifetime of the blue whirl, it transitions several times from the short-lived transitional whirl and then back to the more stable blue whirl (Fig. 4 and Movies S1 and S2). In the transition process, the upper region is replaced by a yellow flame, as shown in Fig. 4*C-O*. Then, the blue cup forms again and holds the yellow flame in its center (Fig. 4*P* and *Q*). Subsequently, the yellow flame spirally shrinks in the blue cup and disappears in the center region of the blue whirl (Fig. 4*R-Y*), leaving a fully-recovered blue whirl. Interestingly, a yellow spiral flame can be observed to be enclosed in an envelope composed of a blue whirl and a secondary violet, conical flame (see Fig. 4*R-V*).

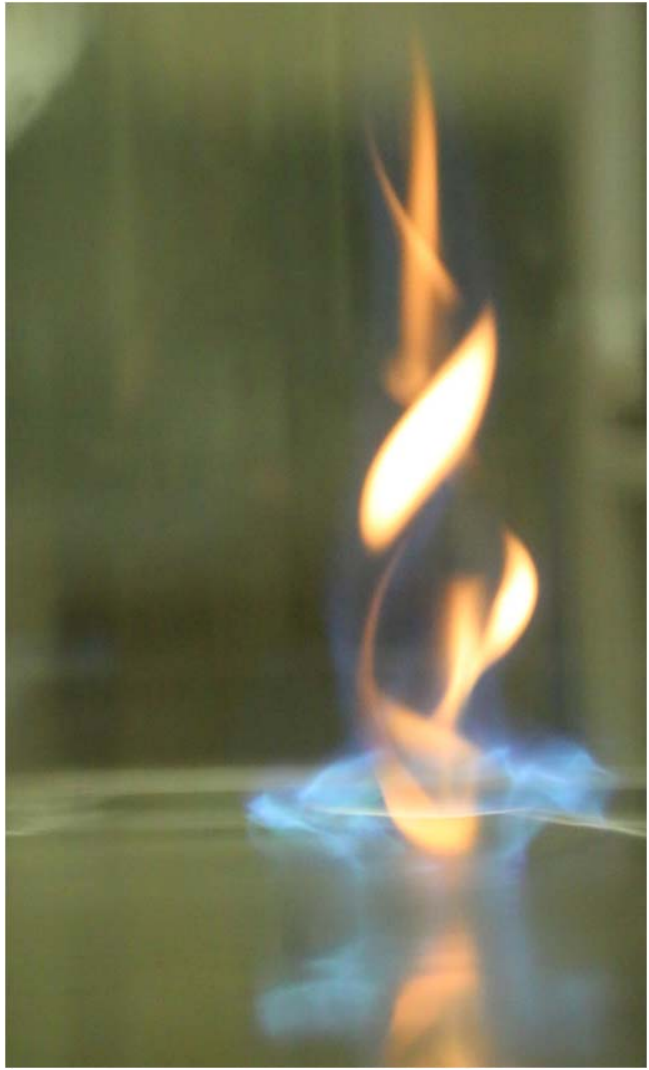

**Figure 3.** Transitioning whirl during the evolution process from a yellow fire whirl to a blue whirl. The height of this structure is about 15 cm. A cup-like blue whirling flame on water holds a yellow whirling flame.

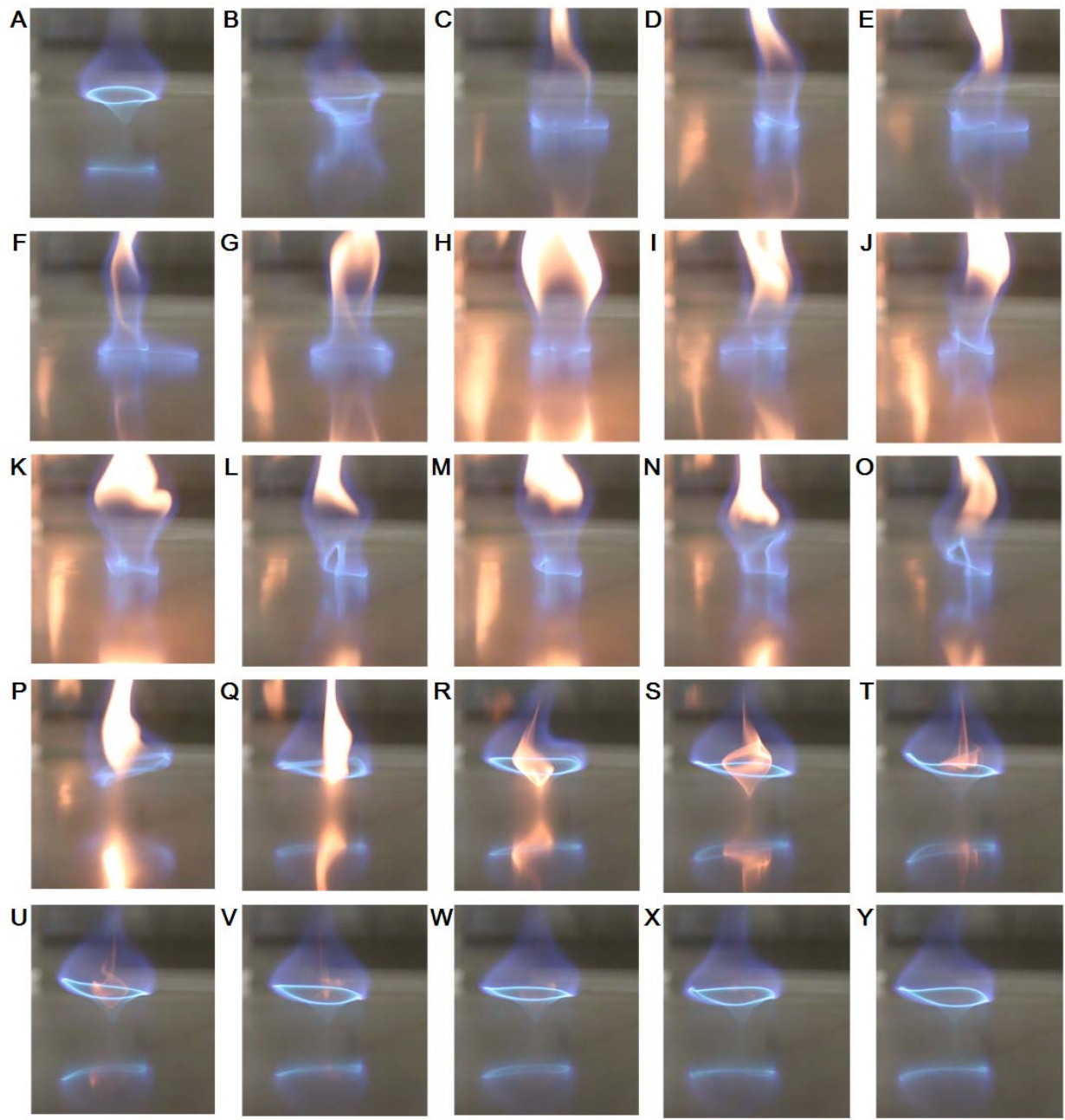

**Figure 4.** Successive frames showing transition between a laminar blue whirl and a yellow whirl taken from a high-speed video. (*A*) Blue whirl. (*B*) Collapse of the blue whirl just before it transitions to a yellow fire. (*C-O*) Transitional yellow fire with a blue base. (*P-X*) Transitional blue whirl holding a yellow whirl in its center. (*Y*) Fully-recovered blue whirl. The entire transition (blue whirl → yellow whirl → blue whirl) takes approximately two seconds. Inside the yellow center of the blue whirl, there appears to be a rotational core that fades as the whirl becomes stabilized over the fuel and a steady regime commences (see Fig. 4*R-U*).

**Understanding the physics of the blue whirl**

We speculate that the boundary conditions, that is, the existence of an air-water-fuel boundary layer instead of a fuel-air boundary on solid surface, plays an important role on both in the increased efficiency of fire whirls on water and the possibility of the transition to a blue whirl. For the fire whirl on a solid surface, there is a flame vortex tube that ends on the solid surface. On water, the flame appears to sit above the surface and even generates vortical motions in and on the water. There is likely to be a layer of evaporated fuel between the bottom of the blue whirl and the water, possibly creating a premixed region at the base. On the other hand, flow rotation intensifies air entrainment and causes strong inflow of air near the boundary layer in vortex phenomena (2, 3, 7, 10), which can result in fast mixing of reactants above the fuel surface. This would have the general form of a triple flame (22), a small premixed flame connected to a diffusion flame. This may also result in higher temperatures and thus increased burning efficiencies.

The yellow color of hydrocarbon diffusion flames, such as pool fires or fire whirls, is due to black-body emissions by radiating soot particles. Soot forms when there is not enough oxygen present to burn the fuel completely. The blueish and violet colors are due to chemiluminescence of excited species such as $C_2$, CH and OH radicals (23, 24). Blue in the whirl indicates that there is enough oxygen present for complete combustion, and therefore suggests a premixed flame. Previous work has shown that fast mixing, which occurs in certain coflowing or opposed-jet diffusion flames, can create soot-free, blue flames (25-27). The explanation was that the fluid dynamics around the flame helped limit soot precursors and soot in fuel-rich regions, so that complete oxidation occurred in fuel-lean regions (26-28). Similar processes may limit the soot formation here.

Fuel rotation induced by the spinning helps the flame stay centered in the tank. The strong rotation lowers the pressure in the center of the vortex, thus keeping the fuel slick from spreading as it does in pool fires. This helps the blue whirl to burn for extended periods, sometimes removing any visible residual fuel slick. This may be useful for *in situ* burning, because an oil slick must be maintained with a critical thickness or the fire will burn out (29).

In conclusion, experiments have discovered a new flame phenomenon, the blue whirl, which shows a distinct flame structure, reduced emissions, and possibly higher combustion efficiencies than both traditional pool fires and fire whirls. These changes are thought to occur due to fast mixing, which limits soot formation in the vortex and creates regions of premixed fuel and oxidizer. Many questions still remain, such as w*hy has this blue whirl not been seen before, what are the physics controlling the formation of the blue whirl, and can we generate blue whirls at larger scales?* Further understanding of the complex, multiphase physics occurring during blue whirl combustion offers dynamic and exciting possibilities for the future, and may therefore lead to the development of novel methods for fuel-spill remediation and high-efficiency combustion.

**Materials and Methods**

The experimental setup is shown in Fig. 5. The experiment was performed on top of a round steel pan filled with water, with a 40 cm inner diameter and 3.2 cm in height. The water surface was flush with the edge of the water pan. Two quartz half-cylinders (30 cm in diameter and 60 cm in height) were suspended about 2 mm over the water surface (Note that the blue whirl can also form when the bottoms of the half-cylinders fully or partly touch the water surface). The ambient pressure and temperature in the experiments were 1 atm and 298 K, respectively. The liquid fuel used was 99.4% pure liquid n-heptane. Initially, the water was quiescent and 2.5 ml heptane was squirted from a syringe onto the center region of the water surface. Shortly thereafter, the heptane was ignited by a small ignitor filled with butane (Olympian GM-3X) which was removed immediately after ignition. The gap size between the half-cylinders and the initial amount of liquid n-heptane poured on the water were initially varied, but the observations reported here were all taken at 1.8 cm and 2.5 ml, respectively.

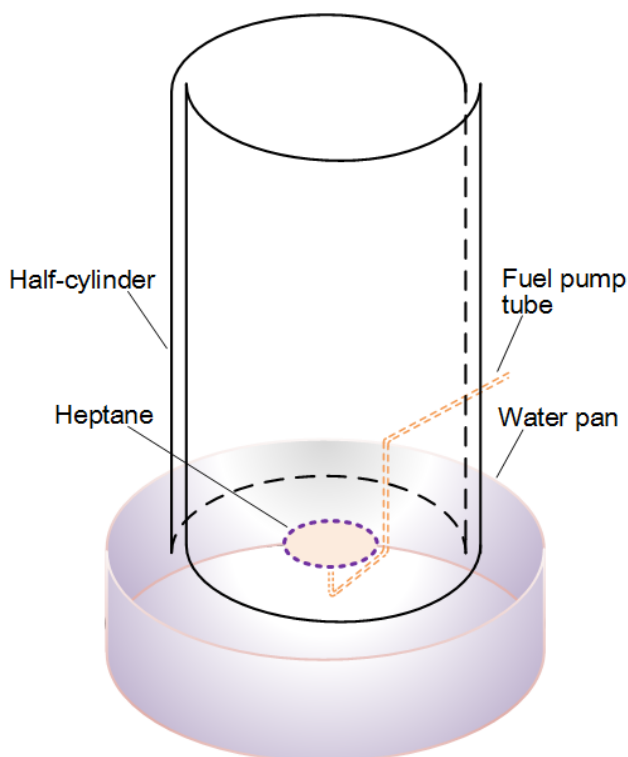

**Figure 5.** Experimental design of the fire whirl generator over water. The copper fuel pump tube, which is connected to a syringe pump, is removable.

To sustain the blue whirl for extended observation, a syringe (BD 60 ml Syringe) was first filled with heptane and then injected slowly to the center region of the water surface through a copper tube (external and internal diameters 0.3 and 0.1 cm, respectively). The copper tube was extended along the bottom of the water pan to the center and bent vertically upward until its opening was about 3 mm below the water surface. A small needle was projected out of the tube opening with its tip just under the water surface to prevent bubble formation when feeding heptane. The copper tube was connected to the syringe using a rubber tube outside the water pan. The injection of fuel was controlled by a syringe pump (Harvard Apparatus Pump 11 Elite) to continually supply heptane at a constant rate (1.1 ml/min in this report).

Changing the exterior shape of the confining apparatus does not seem to affect whether or not fire whirls or blue whirls can form. Experiments in a four-sided square polymethyl methacrylate (PMMA) apparatus produced a fire whirl evolving into a blue whirl. The gap size of the slits between the two half-cylinders, however, does affect the formation and stability of both the blue and traditional fire whirls. Ultimately, a two-gap configuration with a cylindrical shape was chosen for the study in the belief that the cylindrical shape would allow the blue whirl to form and remain stable in the center region of the apparatus. It is worth noting that a blue whirl was even seen to form in tests using heavy hydrocarbons such as crude oil.

The evolution of the fire and fire whirls was recorded using a DSLR camera, a Canon EOS 70D. The images in Fig. 1*A* and *B* were taken using a Tv mode with manual focusing and automatic ISO. The images in Figs. 1*C*, 2 and 3 were taken using Scene (SCN) mode with manual focusing. The images taken by the camera in Figs. 1, 2 and 3 were acquired with a pixel resolution of $3648 \times 3432$ and bit depth of 24. The video from which Fig. 4 was extracted and the videos in Movies S1 and S2 were taken at a frame rate of 50 fps with a pixel resolution of $1280 \times 720$.

**Author Contributions**
H.X., M.J.G. and E.S.O. designed the research; H.X. and M.J.G. carried out the experiments and H.X., M.J.G. and E.S.O. interpreted results and prepared the manuscript.

**Acknowledgments**
The authors would like to acknowledge Ajay V. Singh, Evan Sluder and Sriram Bharath Hariharan for their assistance in laboratory experiments. This work was supported by the National Science Foundation through an EAGER award CBET-1507623 and by the University of Maryland through Minta Martin Endowment Funds in the Department of Aerospace Engineering and the Glenn L. Martin Institute Chaired Professorship at the A. James Clark School of Engineering.